# Understanding the effect of porosity on the polarisation-field response of ferroelectric materials

Yan Zhang[a], James Roscow[a], Rhodri Lewis[b], Hamideh Khanbareh[a], Vitaly Yu. Topolov[c], Mengying Xie[a], Chris R. Bowen[a, *]

[a] *Materials and Structures Centre, Department of Mechanical Engineering, University of Bath, BA2 7AY, United Kingdom*

[b] *Renishaw Plc., Wotton Road, Charfield, Wotton-under-Edge, GL12 8SP, United Kingdom*

[c] *Department of Physics, Southern Federal University, 5 Zorge Street, 344090 Rostov-on-Don, Russia*

**Abstract**

This paper combines experimental and modelling studies to provide a detailed examination of the influence of porosity volume fraction and morphology on the polarisation-electric field response of ferroelectric materials. The broadening of the electric field distribution and a decrease in the electric field experienced by the ferroelectric ceramic medium due to the presence of low-permittivity pores is examined and its implications on the shape of the hysteresis loop, remnant polarisation and coercive field is discussed. The variation of coercive field with porosity level is seen to be complex and is attributed to two competing mechanisms where at high porosity levels the influence of the broadening of the electric field distribution dominates, while at low porosity levels an increase in the compliance of the matrix is more important. This new approach to understanding these materials enables the seemingly conflicting observations in the existing literature to be clarified and provides an effective approach to interpret the influence of pore fraction and morphology on the polarisation behaviour of ferroelectrics. A new general rule to describe the relationship between the polarisation and porosity is also proposed. Such information provides new insights in the interpretation of the physical properties of porous ferroelectric materials to inform future effort in the design of ferroelectric materials for piezoelectric sensor, actuator, energy harvesting, and transducer applications.

**Keywords**:

Porosity, Ferroelectric, External electric field, Coercive field, Morphology

# 1. Introduction

Ferroelectric materials are an important class of electro-active dielectrics that exhibit high levels of polarisation, high permittivity, large electromechanical coupling, and a number of functional properties [1] such as piezoelectric, pyroelectric, electrostrictive, electro-caloric and electro-optic properties for applications related to sensors, actuators, energy harvesters, and transducers [2, 3]. A small amount of porosity is present in all ferroelectric ceramics and is usually considered a defect introduced due to processing imperfections; however, it may also be introduced deliberately to create a composite material system using porosity as a second component [4-6] to provide beneficial effective properties for specific applications. As a result, a thorough understanding of the influence of porosity on the properties of ferroelectric materials remains an important topic. To date, porosity in ferroelectric and piezoelectric materials has been considered at the micro-scale [7] and nano-scale levels [8], including bulk materials [9-12], textured materials[13], aerogels [14] and thin films [8, 15].

One reason for deliberately introducing porosity into a bulk material is to reduce its acoustic impedance, $Z= (\rho Y)^{0.5}$, where $\rho$ is material density, and $Y$ is the Young's modulus of the material. A reduced acoustic impedance improves matching between the piezoelectric material and the fluid, biological medium or material under evaluation when used sensor or transmitter in SONAR, medical ultrasound or non-destructive testing applications [12]. Porosity has also been added to ferroelectric ceramics and/or composites in an effort to reduce the permittivity of the material. The reduction of permittivity leads to an increase in performance for applications related to piezoelectric sensing, pyroelectric sensing and energy harvesting since a variety of performance figures of merit are inversely proportional to the material permittivity at constant stress ($\varepsilon_{33}^T$) [16, 17]. Examples include a variety piezoelectric voltage coefficients ($g_{ij}$), such as $g_{33} = d_{33}/\varepsilon_{33}^T$, pyroelectric sensitivity, $F_v = p/c_E \cdot \varepsilon_{33}^T$, along with associated harvesting figures of merit [7], such as FOM= $d_{33}^2/\varepsilon_{33}^T$; where $d_{33}$ is the longitudinal piezoelectric charge coefficient, $p$ is the pyroelectric coefficient and $c_E$ is the specific heat. The inclusion of pores also results in an increase in the difference between the longitudinal ($d_{33}$) and transverse ($d_{31}$) piezoelectric coefficients which gives rise to improved hydrostatic performance, such as $g_h = (d_{33} + 2d_{31})/\varepsilon_{33}^T$ for a transversely isotropic medium [2, 4]. The possible influence of porosity on the fatigue properties of ferroelectric materials during multiple polarisation reversals has also been considered, for example, for ferroelectric memory and high-strain actuator applications [18]. Piezoelectric surface charges are also thought to influence biological cell attachment and growth, and the fabrication of porous piezoelectric composites with well-defined interconnected pores has been considered to provide a favourable environment for bone ingrowth and osseo-integration [19]. In addition to ceramic systems, porous ferroelectric polymers are also of interest as the presence of poled pores within a polymer matrix can lead to ferroelectret effects [20]. The influence of porosity on the physical properties such as dielectric permittivity and elastic stiffness are relatively well-known, and a number of models that allow prediction of the piezoelectric coefficients are available [5, 6, 17, 21, 22]. However, these models are generally based on the assumption that ferroelectric regions in the porous-ceramic mixture are fully poled in a specific direction. Ferroelectrics are poled by applying an electric field, $E_f$, that is greater than the coercive field of the material, $E_c$, to align ferroelectric domains in grains and achieve a remnant polarisation in the bulk material. In a dense bulk ferroelectric material, the electric field distribution is relatively homogenous [12], leading to a fully poled structure when $E_f > E_c$. However, when pores with a significantly lower permittivity than the bulk ferroelectric are introduced into the microstructure, the applied electric field concentrates in the low-permittivity pore region, which leads to an inhomogeneous electric field distribution throughout the structure [12, 23-26]. The fields are reported to be particularly inhomogeneous close to pore-ceramic interfaces, where low-field regions exist on boundaries aligned parallel to the poling field $E_f$ vector, and high-field regions occur perpendicular to the $E_f$ vector [27]. The inhomogeneous field distribution has led to interest in using porous ferroelectric ceramics as materials with a high degree of tunability by exploiting nonlinear and electric field dependent dielectric properties [11, 23, 27-29].

While there are numerous publications related to the influence of porosity on the stiffness, permittivity, pyroelectric and piezoelectric coefficients of ferroelectric materials [4, 6, 10, 22], less work has examined the effect of porosity on the overall polarisation - electric field (P-E) behaviour, including the shape of the hysteresis loop, remnant polarisation and coercive field. An understanding of this topic is important since the level of polarisation will ultimately affect the final piezoelectric and pyroelectric properties of any porous ferroelectric. Work to date on understanding the polarisation-field response of porous ferroelectrics has considered modelling [23, 30] and experimental approaches [8-10, 15, 31-34], along with combinations thereof [2, 7, 11, 12, 24, 25, 27, 28, 35], which will now be discussed in detail.

**1.1 Impact of porosity on polarisation – field behaviour**

The current literature on the impact of porosity on the polarisation-field characteristic of ferroelectric materials is rather mixed and often contradictory, at both modelling and experimental levels. A typical ferroelectric polarisation - electric field loop contains a saturation polarisation ($P_s$), remnant polarisation ($P_r$) and zero polarisation that is related to the coercive field ($E_c$). It is generally agreed that the porosity reduces both $P_s$ [11, 12] and $P_r$ [2, 7, 10, 12, 14, 15, 32-34, 36-38], and one reason is that a porous ferroelectric ceramic has a lower amount of the active component compared to the dense counterpart and is therefore expected to have a smaller level of polarisation [11]. Nagata [31] showed that experimental data for the remnant polarisation of a porous material reduced more quickly than predicted by a simple volume-fraction relationship given by,

$$P_r = P_r^0 (1-v_p) \qquad \text{(Eqn. 1)}$$

where $P_r^0$ is the remnant polarisation of the dense material, and $v_p$ is the porosity volume fraction. This indicated that there was an additional depolarisation factor due to the presence of the pores, rather than simply a smaller volume fraction of the active ferroelectric component.

An understanding of the impact of porosity on the coercive field of porous ferroelectric materials is less clear. Banno [30] reported that modelling showed that there is no dependence of coercive field on porosity, and this was in agreement with experimental measurements by Nagata on lead lanthanum zirconate titanate (PLZT) materials [31]; the experimental results actually indicated a small increase in $E_c$ with increasing porosity. Zhang et al. [32] experimentally observed either a small or no decrease in coercive field with increasing porosity for a porous lead zirconate titanate (PZT), depending on the pore-forming agent and Li et al. observed no change in $E_c$ with porosity for PZT [37]. This insensitivity of the coercive field with porosity was thought to be due to space charge accumulation at the pore surface. Geis et al. [14] observed a decrease in coercive field with increasing pore volume fraction for porous PZT aerogels when compared to the dense material. The slope of the polarisation-field loop, or degree of rectangularity, was higher for the dense material, and this was considered to be due to the inhomogeneous electric field in the porous material leading to domain switching at different applied electric fields. Okazaki et al. [9] examined the influence of porosity on the properties of PLZT ceramics, and in this case an increase in coercive field was observed with increasing porosity; it was stated that this was due to the space charge field increasing with increasing porosity. Barabanova et al. [33] also observed an increase in coercive field with increasing porosity fraction with $E_c$ = 4.4 kV/cm for the dense PZT and $E_c$ = 6 kV/cm for the same material with 25 vol.% porosity. Porous lead free materials were examined by Yap et al. [36] who stated the coercive field remained stable with porosity content, although the experimental data indicated a small increase in $E_c$ with porosity, although the potential for silicone oil to infiltrate pores during testing was highlighted. Work by Bacaric et al. [10] on porous PZT indicated that the introduction of 30 vol.% of fine-scale pores reduced the $E_c$ value to 6.8 kV/cm compared to the dense material ($E_c$ = 7.8 kV/cm), while larger pores at the same volume fraction led to an increase in $E_c$ to 8.1 kV/cm. The influence of pore geometry on the coercive field of PZT materials was also investigated by Khachaturyan et al. [2]. The experimentally measured coercive field of the dense PZT material ($E_c$ =14 kV/cm) was lower than a porous PZT with anisometric (elliptical) pores ($E_c$ =16 kV/cm), but higher compared to a material with isometric (equi-axed) pores (11 kV/cm). Finite element simulations were used to indicate that the domain switching times were smeared out by introducing porosity into the material structure. The broadening of the switching times was greatest for elliptical pores that were perpendicular to the applied field. Nb-doped PZT materials were examined by Stoleriu et al. [34, 35] and porosity led to the polarisation-field loops becoming less rectangular with more distributed coercivities. First-order reversal curves (FORC) were used to show how dipolar units are spread to higher and lower fields for a porous material, indicating an inhomogeneous ferroelectric system. The dense material exhibited more homogeneous switching properties, since domains are subjected to similar local applied electric fields. It was also thought that the presence of pores relax the level of mechanical clamping[35].

In the majority of the aforementioned bulk ferroelectric materials, porosity is at the micro-scale level. For nanoporous lead titanate thin films, Castro et al. [8] observed that the presence of porosity reduced the coercive field compared to the dense thin film and the presence of nano-scale porosity increased the switching ability. This was thought to be due to porosity triggering the crystallisation of the film at lower temperatures and reduced constraining effect of the substrate on the more compliant porous films. Stancu et al. [15] observed a higher coercive field for the dense PZT thin film compared to the porous films for pore fractions up to 30%, although the dependency with porosity was not clear. The switching characteristics of the porous film was said to be more gradual with applied field and it was thought that the porosity led to different orientations of crystallites within the film.

The most detailed studies of the effect of porosity on ferroelectric and dielectric properties have been examined by Padurariu et al. [11, 23, 27, 29]. The motivation was to exploit the inhomogeneous electric field to provide high tunability. Finite element modelling of porous materials based on lead zirconate titanate niobate (PZTN) based materials indicated that a large component of the electric field was concentrated in the low permittivity pore space and the bulk ceramic is subjected to electrical fields that are very different to the applied external field [11]. Experimental measurements carried out on PZTN indicated that porosity reduced the hysteresis area and a reduction in the rectangularity of the loop ($P_r/P_s$), which relates to tilting of the polarisation field-loop as a result of a broadening of the field distribution. The experimental work indicated that the introduction of approximately 45 vol.% porosity reduced the coercive field compared to the dense material, and the exact level depended on the connectivity of the porous samples; 0-3 or 3-3 connectivity as defined in [22]. A detailed study on Nb-doped PZT ceramics was undertaken by Gheorghiu et al. [12] using experimental characterisation and finite element modelling combined with Preisach modelling to predict polarisation-field loops. Experimental measurements indicated more tilted loops on increasing the porosity from 5 vol.% to 30 vol.%, along with a reduction in rectangularity. The Preisach modelling indicated that higher electric fields were needed to switch non-180° domains for the more porous material due to the high field in the pore space and the resulting lower electric field in the ferroelectric component; however, the experimental results showed that the dense material exhibited a higher coercive field compared to the porous materials.

Clearly there are a number of conflicting and contradictory observations from both experimental and modelling of the polarisation-field response of porous ferroelectrics. The aim of this paper is to provide new insights and a definitive understanding of the role of porosity on the polarisation-field response of ferroelectric materials. This will be achieved by a combination of finite element modelling and careful experimental investigation to understand the influence of porosity volume fraction and morphology on electric field distribution and level of polarisation of a range of porous ferroelectric ceramics with controlled levels and shape. In this work, a lead-free material, barium calcium zirconate titanate whose chemical formula is $0.5Ba(Ca_{0.8}Zr_{0.2})O_3$-$0.5(Ba_{0.7}Ca_{0.3})TiO_3$, hereafter abbreviated as BCZT, was utilised to explore the effect of porosity on the polarisation-field response experimentally, and corresponding modelling work is also presented. One reason for selecting a lead-free ferroelectric system is that it mitigates lead-loss during the high temperature sintering stage, which may especially important for highly porous and high surface area PZT-based ceramics.

## 2. Manufacture and characterisation of equi-axed and aligned porosity

To examine the influence of pore fraction on the shape and key parameters of the polarisation-field loop, the first stage is to manufacture ferroelectric ceramics with either equi-axed or aligned porosity at a wide range of volume fractions, up to 40 vol.% porosity. The BCZT powders were synthesised by the traditional solid-state reaction method using $BaCO_3$ (99+%, Acros Organics), $CaCO_3$ (99.9%, Sigma Aldrich), $TiO_2$ (99.9%, rutile, Sigma Aldrich) and $ZrO_2$ (99%, Sigma Aldrich). The raw materials were weighed according to the stoichiometric formula. The mixed powders were then ball-milled with ethanol and $ZrO_2$ balls for 4 h. Thereafter, the obtained mixtures were dried and calcined at 1200 °C for 3 h and followed by additional ball-milling for 24 h.

To fabricate ceramics with randomly distributed equi-axed pores the burned out polymer spheres (BURPS) process was used where the milled powders were mixed for 12h with ethanol with different weight fraction additions of a pore forming agent, polyethylene glycol (PEG) at weight fractions of 0, 3, 5, 9, 13, 16, 20, and 23 wt.% based on the calcined powders. This was followed by mixing with 1 wt.% polyvinyl alcohol (PVA) binder to aid cold uniaxial pressing. After drying, the powders were uni-axially pressed to form pellets of 13 mm in diameter and 1.5 mm in thickness. The sample with 0 wt.% PEG addition is the 'dense' material that will be used as a reference for comparison with the porous BCZT.

To generate aligned porosity, freeze casting was carried out by pouring water-based BCZT suspensions with different levels of BCZT solid loading levels into a transparent cylindrical polydimethylsiloxane (PDMS) mould and subjecting the mixture to unidirectional freezing. This was followed by a freeze drying process reported in our previous studies [7, 39-41] to remove the ice crystals and produce aligned BCZT powder sample. Water was utilised as the solvent to disperse the ceramic for freeze casting, which has been widely explored to obtain long-range ordered anisotropic aligned pore channels [42].

To sinter the porous materials made by both BURPS and freeze casting, the pressed BCZT pellets and freeze-dried green bodies were initially heated to 600 °C for 3h to remove organic additives and then sintered at 1400 °C for 4 h in air to achieve the final sintered samples. The bulk densities of the sintered materials were measured using the Archimedes' principle. For evaluation of the electrical properties, silver paint was coated on both faces of the sintered samples to form electrodes. The saturation polarisation, remnant polarization and coercive field were measured using a Radiant RT66B-HVi Ferroelectric Test system on initially unpoled materials to ensure no impact of a previous poling process on the P-E loop with the hysteresis period of 200 ms. The samples were tested in air to mitigate any impact of oil in the pores on the P-E response, as highlighted by Yap et al. [36]. For the modelling activities the relative permittivity of the unpoled and poled material was measured using a Solartron 1260 and 1296 Dielectric Interface.

## 3. Experimental results

### 3.1 Equi-axed and randomly distributed porosity

Figure 1 (A-H) shows scanning electron micrographs of BCZT with different porosities from 4-40 vol.%, which correspond to PEG additions from 0 to 23 wt.%. The micrograph in Figure 1(A) is related to a BCZT ceramic with no pore-forming agent addition, and this sample is dense with a measured porosity of ca. 4 vol.%.

It can also be seen that after the addition of PEG pore-former from 3 wt.% to 23 wt.%, the number of the pores in the ceramic matrix increased, and a microstructure with a fully densified polycrystalline matrix can be seen for all porous ferroelectrics in Figure 1(B-H). These morphologies indicate that the BCZT powders were well sintered under the processing condition and the microstructures for all levels of porosity exhibited homogeneously distributed and equi-axed spherical pores in the ceramic matrix. Pore size and interconnection was found to increase with increasing fraction of pore forming agent as neighbouring pores tend to coalesce, e.g. the pore size of 50 μm for the porous ceramic with 10 vol.% porosity shown in Figure 1(B), compared with the pore size with approximately 100 μm when the porosity volume fraction was 25 % shown in Figure 1(E). It was also seen that a continuous porous network was formed in the porous BCZT ceramics and the homogeneous pore distribution led to the formation of a strong pore wall network. By control of the weight fraction of the PEG pore-former, a range of porous ceramics was readily achieved. The pore network, electromechanical properties and modelling of porous PZT-type materials have also been described by Filippov et al. [43] in the porosity range from 5 to 57 vol. % with similar porous structures observed in the PZT-type [43] when compared to the BCZT samples in Figure 1, although the impact of the pores on the polarisation-field behaviour was not studied.

Figure 2(A-C) shows the effect of porosity volume fraction on the P-E loop, remnant polarisation and coercive field for the dense material (4 vol.% porosity) and the porous BCZT ceramics with randomly distributed equi-axed porosity ranging from 10 vol.% to 40 vol.%. As can be seen from Figure 2(A), the remnant polarisation ($P_r$) gradually decreased with an increase in the fraction of porosity. The dense material had a $P_r$ of 7.7 μC/cm$^2$, which compares favourably with previously reported values [36, 44, 45] that range from 6.3-8.0 μC/cm$^2$. As the porosity volume fraction is increased, the reduced fraction of the ferroelectric ceramic leads to a reduction in $P_r$ from 6.2 μC/cm$^2$ at $v_p$ = 10 vol.% to 1.3 μC/cm$^2$ at $v_p$ = 40 vol.%, as seen in Figure 2(B). A decrease in $P_r$ with increasing porosity has been observed elsewhere [2, 10, 12, 32, 36] and it can been seen in Figure 2(B) that the experimentally measured $P_r$ decreases at a faster rate compared to the volume rule of mixtures given by Eqn. 1, similar to that observed by Nagata [31] and Gheorghiu [12]. This indicates that the reduction in the fraction of ceramic is not the only reason for the decreased $P_r$ and there is an additional depolarisation factor due to the presence of pores, which will be discussed in more detail in relation to the modelling results described later.

In contrast to the continually decreasing $P_r$ with increasing porosity level, Figure 2(C) shows that the coercive field initially decreased from 2.36 kV/cm to 1.94 kV/cm as the porosity volume fraction increased from 4 vol.% to 25 vol.%; with an almost unchanged rectangularity ($P_r/P_s \approx 0.79$). As the porosity level increased from 25 vol.% to 40 vol.% the coercive field then increased from 1.94 kV/cm to 3.39 kV/cm with a relatively sharp decrease in rectangularity from 0.79 to 0.65. The increased tilting of the polarisation-field loops with increasing pore volume fraction in Figure 2(A) and reduced rectangularity at porosity fractions over 25% in Figure 2(C) is indicative of domain switching occurring over a wider range of electric field levels in the higher porosity materials, compared to the more dense materials.

**3.2 Aligned porosity**

Figure 3(A-D) examines the effect of porosity volume fraction on the polarisation-field behaviour, remnant polarisation and coercive field of dense BCZT (with a porosity of 4 vol.%) and the porous materials formed by freeze casting with aligned porosities that range from 10 vol.% to 40 vol.%. The morphology of the pore channels obtained with different porosities are similar to previous observations for a variety of other ferroelectrics and ceramics dispersed in water-based suspension during freeze casting, such as hydroxyapatite[41], PZT[7], and hydroxyapatite/barium titanate composite[19]. As an example, the microstructure of the porous freeze-cast BCZT with a porosity volume fraction of 40 vol.% is shown in Figure 3(A), where the arrow indicates the freezing direction, along which the pore channels and ceramic are aligned. In this case the polarisation-field loops in Figure 3(B) are all measured parallel to the freezing direction.

The porosity dependence of the remnant polarisation for materials with aligned pore channels has some similarities to the material with randomly distributed equi-axed pores, as seen by comparing Figure 2(B) and Figure 3(C). The $P_r$ gradually decreased from 6.5 to 3.1 µC/cm$^2$ as the porosity fraction increased up to 40 vol.%, however it can be seen that the $P_r$ of the freeze-cast materials decreased more slowly with an increase in pore fraction compared to materials with equi-axed porosity, and was therefore closer to that the $P_r$ value predicted by Eqn. 1, demonstrating a reduced additional depolarisation factor due to the presence of aligned porosity.

Figure 3(D) shows that the $E_c$ decreased to a small extent from 2.3 to 2.0 kV/cm as the porosity fraction increased from 4 vol.% to 20 vol.%. At the higher porosity fractions, there was again an increase in $E_c$ from 2.0 to 3.5 kV/cm as the porosity increased to 40 vol.%, with a corresponding decrease in rectangularity from 0.80 to 0.72 and increased tilting of the polarisation-field loops in Figure 3(B). On comparing Figure 2(C) and Figure 3(D), it can be seen that the $E_c$ and rectangularity of the material with the aligned porosity is less sensitive to the porosity volume fraction compared to the materials with equi-axed pores.

To summarise, the main experimental observations are as follows:

(i) $P_r$ decreases with increasing pore fraction and at a rate that is more rapid than Eqn. 1, indicating that there is an additional depolarisation factor and it is not simply related to a reduction in the volume fraction of the ferroelectric material,

(ii) $P_r$ falls at a more rapid rate with porosity for materials with randomly distributed and equi-axed pores (a decrease from 6.2 to 1.3 µC/cm$^2$ as porosity increases from 10 to 40 vol.%) compared to materials with pores aligned along the polarisation direction (a decrease from 6.5 to 3.1 µC/cm$^2$ as porosity increases from 10 to 40 vol.%),

(iii) the presence of porosity leads to tilting of the polarisation-field loop, in particular at high pore volume fractions when $v_p \gtrsim$ 20-25 vol.%,

(iv) when $v_p \gtrsim$ 20-25 vol.% the introduction of porosity leads to an increase in $E_c$ and a reduction in rectangularity for materials with increasing porosity fraction,

(v) when $v_p \gtrsim$ 20-25 vol.% materials with equi-axed pores exhibit a larger increase in $E_c$ and greater decrease in rectangularity compared to materials with aligned pores,

(vi) when $v_p \lesssim$ 20-25 vol.% the introduction of porosity leads to a reduction in $E_c$ with increasing porosity fraction.

**4. Finite element modelling of polarisation-field loops in porous ferroelectrics**

To improve our understanding of the experimental results, finite element modelling of the electric field distribution in porous ferroelectric materials is now undertaken. This aims to study the impact of the electric field distribution on the polarisation-field response. The effect of both randomly distributed equi-axed pores, such as those in Figure 1, and aligned pores based on the structures produced by the freeze casting method, as in Figure 3A, will be investigated.

A finite element mesh was initially formed that consisted of 40 x 40 x 40 (i.e. 64,000) cubic elements. Elements were assigned the properties of either unpoled BCZT with a relative permittivity of $\varepsilon_r = 1192$ that was measured experimentally, or air with $\varepsilon_r = 1$. To create a model that represented randomly distributed and equi-axed porosity, elements were selected at random in the model and associated with the properties of air until the desired volume fraction of porosity had been achieved; the remaining elements had the properties of BCZT. An example of a model with equi-axed porosity is shown in Figure 4(A).

To create a model consisting of aligned porosity that were typical of freeze casting, pore channels were initially created to form an ideal 2-2 type composite structure [22] that consisted of layers of ceramic or air. Since freeze cast structures have some interconnection between the pore channels, see Figure 3(A), some ceramic elements were introduced at random in the pore channels and a small fraction of porosity (air) was introduced into the ceramic channels, see Figure 4(B) for an example of an aligned model. The voltages at the top and bottom surfaces of the model were coupled to simulate electrodes.

Polarisation-electric field (P-E) loops were simulated by applying a voltage profile similar to those applied in the experimental measurements, shown in Figure 4(C), to an initially unpoled material and sweeping the polarity to positive and negative direction. After each step, the electric field in each BCZT element in the finite model was analysed and when the local field in the z-direction (i.e. the poling axis) was greater than the coercive field of BCZT for the dense material ($E_c$ = 2.4 kV/cm from experiment in Figure 2(C)) the material properties of the element were changed from unpoled BCZT to that of poled BCZT ($\varepsilon_r$ = 2590, determined experimentally). The model took into account only two possible poling directions for the BCZT sample, namely +z and –z that was parallel to the applied electric field; other polarisation directions normal to the applied electric field were initially considered (i.e. +x, -x, +y and -y) but found to have little effect on the results. A schematic of the modelling process to generate P-E loops is shown in Figure 4(D). The model was designed to investigate the effect of porosity on the coercive field of ferroelectric materials due to the inhomogeneous field distributions as a result of the difference in permittivity between the low-permittivity pores and the high permittivity ferroelectric matrix, and the material properties of the ferroelectric component were assumed to be the same as those measured from the dense material across the range of porosities investigated. The change in polarisation was calculated from the measured change in capacitance from the initially unpoled BCZT to the poled BCZT and was normalised with respect to the fully dense material.

**4.1 Finite element model results and discussion**

The modelled polarisation-field loops are shown in Figure 5(A) and (B) for randomly distributed equi-axed pores and aligned porosity, respectively. A gradual reduction in remnant polarisation is observed as porosity in introduced into the structure, along with increased tilting of the polarisation-field loops that is in qualitative agreement with the experimental data in Figures 2(A) and 3(B).

The variation of remnant polarisation with porosity for both equi-axed and aligned porosity in BCZT is shown in Figure 5(C) and experimental data is plotted alongside for comparison, where all data is normalised with respect to the dense BCZT material. Good agreement between the model and experimental data is observed, where the $P_r$ decreases with increasing porosity and at a faster rate for the randomly distributed equi-axed porous structures compared to the aligned structures. The model data also indicates that the remnant polarisation falls more quickly that that predicted by Eqn. 1.

The variation of coercive field with pore volume fraction for the two types of pore structure is shown in Figure 5(D), along with experimental data for comparison. The coercive field of the randomly distributed equi-axed porous BCZT increased with increasing porosity, although not as fast as the experimentally measured data. The rate of increase was higher for the more porous materials. The model data for the aligned porous structure showed very little change in coercive field with increasing porosity, with a slight initial decrease followed by a small increase above 40 vol.% porosity.

In order to understand the influence of porosity on the polarisation-field response it is informative to examine the electric field distribution as a function of pore volume fraction and shape for both the equi-axed pores, Figure 6(A), and aligned pores, Figure 6(B). The plots in Figure 6(A)-(B) indicate the frequency at which different local electric fields exist in the direction of the applied field (z-direction) in the ceramic elements, normalised with respect to the applied field; the frequency of each interval in these plots has been normalised to the total number of ferroelectric elements in each model such that the integral of each curve is one. For simplification, we examine the case when the applied electric field is significantly lower than the coercive field so that there is no switching in the model from properties in the unpoled state to the properties in the poled state.

We now consider the electric field distribution in Figure 6(A) in the model with randomly distributed equi-axed porosity. In the dense material (4 vol.% porosity) the electric field is relatively homogenous with a strong peak slightly below $E_{local} / E_{applied} = 1$. This homogenous field throughout the material leads to a rectangular polarisation field-loop in Figure 5(A), with limited tilting of the loop since poling occurs throughout the material when the applied electric field is equal to the coercive field.  As equi-axed porosity is randomly introduced into the structure, and the $E_{local}/E_{applied}$ peak shifts to lower values of field, see Figure 6(A), so that the mean and mode field in the ceramic is lower than the applied electric field. This occurs since ~~to~~ the electric field is concentrating in the low permittivity pores; Figure S1(A) shows the increased field in the pore elements. Increasing the fraction of porosity results in a further reduction in the observed field in the high permittivity ceramic phase. The distribution of the electric field also broadens, see Figure 6(A), as the porosity volume fraction increases since the field distribution throughout is becoming more complex in the porous structure.

This broadening of the electric field distribution leads to a switching behaviour over a broader range of applied electric fields and tilting of the polarisation-field loop, as seen in both the experimental and model data in Figures 2(A) and 5(A) respectively. This shift in peak electric field to lower values with increasing porosity also explains the increase in coercive field with porosity level predicted by the model, see Figure 5(D), since higher applied fields are required to switch the net polarisation orientation of the porous material compared to the dense material.

Electric field distributions are shown in Figure 6(B) for aligned porosity that has pore channels and ceramic aligned parallel to the direction of applied field, as in Figure 3(A) and Figure 4(B). These porous structures can be seen to have a more homogenous electric field distributions compared to randomly distributed equi-axed pores. In this case, as material is aligned in the direction of applied electric field there is no significant reduction in mean electric field, compared to Figure 6(A), although a small degree of broadening of the field distribution is observed compared to the dense material with 4 vol.% porosity. Figure S1(B) also reveals a less broad field distribution in the pore region, compared to the system of equi-axed pores in Figure S1(A).

To understand why porosity leads to a broadening of the electric field distribution within the material, we now show Figure 6(D) that represents a two-dimensional electric field nodal contour plot of a pore in an unpoled BCZT ceramic medium. While the presence of randomly distributed pores leads to an overall reduction in the mean field in the BCZT medium, Figure 6(A), it can be seen in Figure 6(D) that regions close to the pore edges that are parallel to the direction of the applied field experience electric field concentrations. Yap et al. also observed high and low electric field intensities around pores [36]. This leads to the ferroelectric material in these high field regions becoming poled, or in the case of the P-E loop measurements, undergoing switching at applied fields below the coercive field. Regions at the pore edges that are perpendicular to the applied electric field experience electric fields lower than the applied field, and higher applied electric fields are necessary to pole these regions. The presence of these field distributions leads to tilting of the polarisation-field loops in Figure 5(A), broadening of the field spectra and a reduction of mean field in Figure 6(A), and an increase in coercive field with an increase in porosity in Figure 5(D). The remnant polarisation decreases with increasing porosity at a more rapid rate compared to a simple volume fraction dependency, as in Eqn. 1, since the field distribution and the existence of field concentrations in the low permittivity pore space (Figure S1) both represent an additional depolarisation factor. It should be highlighted that Figure 6(D) is a relatively simple model to understand the electric field distribution around a single pore, while for high porosity volume fractions the interaction of electric fields and pores can occur; as in the larger models of Figure 4.

Increasing the aspect ratio of pores in the direction of the applied field, such as the case in the freeze cast structures, leads to a higher volume of material at pore edges which are parallel to the applied field being subject to high electric fields, which explains the slight decrease in $E_c$ with increasing porosity for the aligned porous structures in Figure 5(D), and why some regions experience local fields greater than the applied field ($E_{local}/E_{applied}$ > 1) in Figure 6(B). This is in Figure 6(C), which is a comparison of the electric field distribution for dense BCZT (4 vol.% porosity) and BCZT for approximately 50 vol.% equi-axed and aligned porosities. A small secondary peak for the aligned porous BCZT is observed that is likely to arise from isolated ceramic elements/bridges in the pore channels. If the pores were aligned perpendicular to the applied electric field, higher $E_c$ values would be expected, as has been observed in freeze-cast PZT materials [7].

It is now possible to explain the majority of the experimental observations, points (i)-(vi), based on the finite element modelling.

(i) $P_r$ decreases with increasing porosity and at a rate that is more rapid than Eqn.1, see Figure 5(C), due to an additional depolarisation mechanism that results from the reduction in electric field in the high permittivity ceramic, Figure 6(A), and a concentration of electric field in the lower permittivity air region (Figure S1).

(ii) $P_r$ falls at a more rapid rate for ferroelectric materials with randomly distributed equi-axed pores compared to materials with pores aligned along the polarisation direction. This is due to a combination of the more complex field distribution in porous materials with randomly distributed equi-axed pores compared to the aligned porous structures, and the greater fraction of pore-ceramic interfaces perpendicular to the applied field that are subject to electric fields far below the applied field, see Figure 6(C and D).

(iii) Broadening of the field spectra due to the presence of porosity leads to tilting of the polarisation-field loop in both experimental and modelled polarisation-field loops, in particular at high pore volume fractions, since poling is taking place over a greater range of applied electric fields, see Figures 2(A), 3(B), 5(A) and 5(B).

(iv) The experimental data indicates the introduction of porosity leads to an increase in $E_c$ when $v_p \gtrsim$ 20-25 vol.% and the modelling indicates that this is due to a reduction in electric field experienced by the ferroelectric. The rate of increase is larger in the experimental data compared to the model, however, the model assumes a discrete switching point at $E_c$, whereas in reality a variety of domain types exist in real materials that have a range of switching fields, as in Figure 2(A) for the dense material.

(v) Materials with equi-axed pores exhibit a larger increase in $E_c$ at high porosity volume fractions compared to aligned porosity materials in the modelling and experimental data, see Figure 5(D). This is due to the lower electric field in the equi-axed materials, Figure 6(C).

While the finite element model can be used to interpret the majority of the experimental observations, it cannot explain why for low pore volume fractions, $v_p \lesssim$ 20-25 vol.%, the introduction of porosity leads to a reduction in $E_c$ with increasing porosity, as shown in Figure 2(C) and 3(D).

The porous model discussed here only considers the complex electric field distributions that occur on application of an external electric field and its impact on properties, such as coercive field. Comparing the experimental and the modelled data in Figure 5(D), it appears there are two competing mechanisms that occur when porosity is introduced into these materials. Firstly, the experimental data presented here shows a decrease in the coercive field for relatively low fractions of porosity ($v_p \lesssim$ 20-25 vol.%). Stoleriu et al. [35] indicated that the presence of pores can relax mechanical clamping of the ceramic material. For thin film materials, Johnson-Wilkie et al. [46] describe an increase in ferroelectric response due to the introduction of low levels of porosity, where a reduction in the constraining effect of the substrate at small fractions of porosity can improve domain motion. While electric field enhancement near the pores, as in Figure 6(D), was discussed as a potential mechanism for the decrease in $E_c$ the large changes in ferroelectric properties for relatively low pore fractions was thought to be more likely to be the result of an increase in compliance which facilitates ferro-elastic reorientation. At low porosities ($v_p \lesssim$ 20-25 vol.%) the contribution from the effect of the complex field distribution due to the difference in permittivity between the pore and the ferroelectric component on the coercive field is relatively small with less than a 10% increase in $E_c$ predicted by the porous network model at 25 vol.% porosity. However, when porosity is greater than 30 vol.% the effect on the electric field distribution within the material becomes more pronounced and, therefore, the dominant mechanism, hence the observed increase in coercive field around this porosity. The transition may be influenced by a number of factors, such as the morphology and distribution of the pores, the level of domain mobility (ferroelectric 'hardness') and the elastic compliance of the ceramic matrix or any substrate effects if the material is in a thin-film form.

**5. A re-examination of the literature reporting the coercive field of porous ferroelectrics**

It is now of interest to examine if our understanding of the impact of porosity on the polarisation-field behaviour can be used to address the conflicting reports in the literature. As the decrease in remnant polarisation with porosity level is consistent and relatively clear, the focus is on the coercive field data when contradictory reports of a decreasing or increasing coercive field with increasing porosity are reported. Table 1 and Figure 7(A) summarise the effects of the porosity volume fraction on the $E_c$ value of the porous ferroelectric ceramics from the literature and this work; the data has been separated to those where $v_p \lesssim$ 20-25 vol.% and $v_p \gtrsim$ 20-25 vol.%. It can be seen in Table 1 and Figure 7 that when the porosity volume fraction is low ($v_p \lesssim$ 20-25 vol.%) the $E_c$ tends to decrease with increasing porosity irrespective of whether the pore morphology is random or aligned. The increase in $E_c$ with increasing pore fraction is thought to occur in the higher porosity materials and agrees with our hypothesis that high pore contents lead to a lower magnitude and broader electric field distribution in the ceramic component. Not all systems behave the same and this may be due to the factors such as the size of the pores, the pore size distribution or introduction of defects rather than the controlled pore distributions examined in the porous materials manufactured in this study. It is worth noting that the pore size in the models is relatively small compared to the sample thickness, which is representative of a bulk material. However, for thin films the electric field distribution can be more homogenous if the pores on the same scale as sample thickness. The greatest deviations in Figure 7(A) are related to the work of Liu et al. [47] where calcium was used both as a pore former and dopant to change the ferroelectric properties and Stancu et al. [15] who indicated that for thin films the pores can also influence the electrode/PZT interface properties. Lead-loss in highly porous materials can also have an impact on ferroelectric response and it has been suggested that space charges can cluster around inhomogeneities [9] and can inhibit domain motion in ceramic grains.

Figure 7(B) shows the relationship between porosity and the normalised remnant polarisation $P_r/P_r^0$ for ferroelectric ceramics with equi-axed porosity, based on data reported in the literature. A linear fit for the data is shown, the gradient of which we propose as a 'depolarisation' factor, $d_P$, such that the remnant polarisation at a given porosity can be estimated by

$$P_r = d_P \, P_r^0 \, (1 - v_p) \qquad \text{(Eqn. 2)}$$

This depolarisation factor takes into account the additional volume of high permittivity material surrounding a pore that exists in the low field region and thus contributes to an additional reduction in polarisation that is not considered by Eqn. 1 of Nagata l [31], which is effectively a linear rule of mixtures. For the materials in Figure 7(B) with equi-axed pores, we obtain a depolarisation factor of $d_p \sim 0.6$. The value will be influenced by the pore morphology. For example, for freeze-cast BCZT in Figure 3(C) and PZT where pores are aligned in the electric field direction we obtain $d_P \sim 0.8$, while for freeze cast PZT poled perpendicularly to the pore direction, a value of $d_P < 0.5$ was obtained.

**Conclusions**

This paper has provided a detailed examination and description of the influence of porosity on the polarisation-electric field behaviour of ferroelectric materials. The effects of volume fraction and morphology of pores on the polarisation-electric field loops have been examined both experimentally and by modelling, along with its impact on remnant polarisation and coercive field. The introduction of a low permittivity pore in a high-permittivity ceramic matrix has been shown to lead to a broadening of the electric field distribution in the ceramic component and a decrease in the electric field experienced by the ceramic due to a concentration of the electric field in the lower permittivity pore region. This leads to a reduction in the remnant polarisation $P_r$ and tilting of the polarisation-field loops of the porous material. The variation of coercive field $E_c$ with changes of the porosity level is more complex. The addition of a relatively low volume fraction of porosity ($\lesssim$ 20-25 vol.%) is observed to lead to a decrease in coercive field with increasing porosity, and this is likely to be due to increased compliance of the ceramic matrix which facilitates non-180° domain switching. At higher porosity levels ($\gtrsim$ 20-25 vol.%) the electric field is reduced significantly in the ceramic component so that higher externally applied electric fields are required for switching and the coercive field begins to increase with increasing porosity levels. Equi-axed pores influence the electric field distribution in the ceramic to a greater extent compared to material with pores aligned in the direction of applied field. A new general rule to describe the relationship between the remnant polarisation and the porosity is also proposed with a depolarisation factor ($d_P$) for both equi-axed and aligned porosities.

This work has enabled the seemingly contrasting and conflicting observations in the existing literature to be understood and provides a new approach to understand the influence of pore fraction and morphology on the polarisation behaviour of modern ferroelectric materials. Such information provides new insights in the interpretation of ferroelectric properties of porous materials and can inform future approaches for the design of porous ferroelectrics for sensors, actuators, harvesting and transducer applications.

**Acknowledgements**

Prof. C. R. Bowen and Dr. Mengying Xie would like to acknowledge funding from the European Research Council under the European Union's Seventh Framework Programme (FP/2007-2013) / ERC Grant Agreement No. 320963 on Novel Energy Materials, Engineering Science and Integrated Systems (NEMESIS). Dr. Y. Zhang would like to acknowledge the European Union's Horizon 2020 research and innovation programme under the Marie Skłodowska-Curie Grant with the Agreement No. of 703950 (H2020-MSCA-IF-2015-EF-703950-HEAPPs). Dr. V. Yu. Topolov would like to thank for funding from the Ministry of Education and Science of the Russian Federation (project contract No. 03.G25.31.0276, May 29th, 2017).

**Figure captions**

**Figure 1** Scanning electron micrographs of fracture surfaces of (A) dense (4 vol.%) and porous BCZT with the porosity of (B) 10 vol.%, (C) 15 vol.%, (D) 20 vol.%, (E) 25 vol.%, (F) 30 vol.%, (G) 35 vol.%, (H) 40 vol.%.

**Figure 2** (A) polarisation-field loop for range of $v_p$, (B) remnant polarisation and (C) coercive field, $E_C$, of BCZT with randomly distributed porosity; microstructures are in Figure 1.

**Figure 3** (A) Scanning electron microscopy image of porous freeze-cast BCZT with porosity of 40 vol.%, (B) polarisation-field loops along freezing direction with different pore volume fractions, (C) remnant polarisation and (D) coercive field and rectangularity of freeze-cast BCZT.

**Figure 4** Example model geometry for (A) equi-axed pores (30 vol.% porosity) and (B) aligned pores (33 vol.% porosity), (C) Voltage profile applied to the model to generate polarisation-field loops, and (D) is a schematic of the modelling process.

**Figure 5** Polarisation-field loops for (A) BCZT with random porosity and (B) for BCZT with aligned porosity, and variation of (C) remnant polarisation and (D) coercive field with porosity of both model and experimental data (normalised to the properties of the dense BCZT sample).

**Figure 6** Electric field distributions from modelling with (A) equi-axed porosity (vol.%) distributed at random in the structure, (B) aligned porosity (vol.%) aligned to the direction of the applied field, to simulate the type of structure obtained via the freeze casting process, and (C) a comparison of the spectra for dense BCZT (4 vol.% porosity) and BCZT with ~50 vol.% equi-axed and aligned porosities; field distribution curves have been normalised to the total number of ferroelectric elements in each model, such that the interval under each curve is one. (D) Electric field contour plot from two-dimensional finite element model demonstrating the inhomogeneous field distribution in the vicinity of a single pore.

**Figure 7** (A) Change of coercive field $E_c$ with porosity (PZT film is plotted in inset as $E_c$ is much higher than other materials) and (B) effect of equi-axed porosity on remnant polarisation, $P_r$, normalised with respect to the dense material, $P_r^0$ [2, 7, 9-12, 14, 31, 32, 34, 36, 37, 47].

Table 1 Summary of coercive field $E_c$ in porous ferroelectric materials with porosity volume fraction ($v_p$).

| Porosity ($v_p$) range | Coercive field ($E_c$) trend with increasing porosity | Material | Porosity fractions ($v_p$) | Remarks and pore morphology | Ref. |
|---|---|---|---|---|---|
| $v_p \lesssim$ 20-25 vol. % | Decrease | PZTN (bulk) | Dense, 20 % | Rounded pores | [12] |
| | Decrease | PZT (bulk) | Dense, 30 % | Spherical pores, small pore size | [10] |
| | Decrease | PZT (bulk) | Dense, 17, 25 % | Aerogels | [14] |
| | Decrease | PZT (bulk) | Dense, 20 % | Aligned pores | [7] |
| | Decrease | BCZT (bulk) | Dense, 10, 15, 20 % | Aligned or spherical pores | This work |
| | Decrease | BCZT (bulk) | Dense, 3.9, 20.8, 25.4 % | Spherical pores, elongated and oriented pores. | [36] |
| | Decrease | PZT (bulk) | Dense, 8, 12, 20 % | Irregular pores | [32] |
| | Constant | PZT (bulk) | Dense, 9, 15, 23 % | Spherical pores, applied electric field were not the same for all porosity levels | [32] |
| | Constant | PZT (bulk) | Dense, 16.6 % | Elongated pores | [37] |
| | Constant | PLZT (bulk) | 0.5 to 1 % | Porosities range explored low | [9, 31] |
| | Decrease | PT (thin film) | Dense and porous | Porosity was not mentioned; thickness of the film was 100 nm with the pore size of 50 nm | [8] |
| $V_p \gtrsim$ 20-25 vol. % | Increase | PZT (bulk) | 25 % | - | [33] |
| | Increase | PZT (bulk) | 30 % | Spherical pores, large pore size | [10] |
| | Increase | PZT (bulk) | 30, 40, 50, 60 % | Aligned pores | [7] |
| | Increase | BCZT (bulk) | 25, 30, 35, 40 % | Aligned or spherical pores | This work |
| | Increase | PZT (bulk) | 37 % | Aniso-metric pores | [2] |
| | Decrease | PZT (bulk) | 37 % | Iso-metric pores | [2] |
| | Decrease | PZT (bulk) | 35, 41 % | Irregular pores, applied electric field were not the same for all porosity levels | [32] |
| | Almost constant | PZT (bulk) | 32, 38 % | Spherical pores, applied electric field were not the same for all porosity levels | [32] |
| | Decrease | PZTN (bulk) | 44.5 % | 0-3 and 3-3 connectivity between PZTN and pore (spherical and irregular morphology) | [11] |
| | Decrease | PZT (bulk) | 38, ~41.5, ~46.5, 52.6, 55.5 % | Ca-doping to provide porosity | [47] |
| | Decrease | PZTN (bulk) | 37 % | Nb-doped, lamellar pores | [34] |
| | Increase/Decrease | PZT (film) | 25, 30, 40 % | Nano-scaled pore size from microscopy observation. At 40% highest $E_c$, and at 30% lowest $E_c$ | [15] |

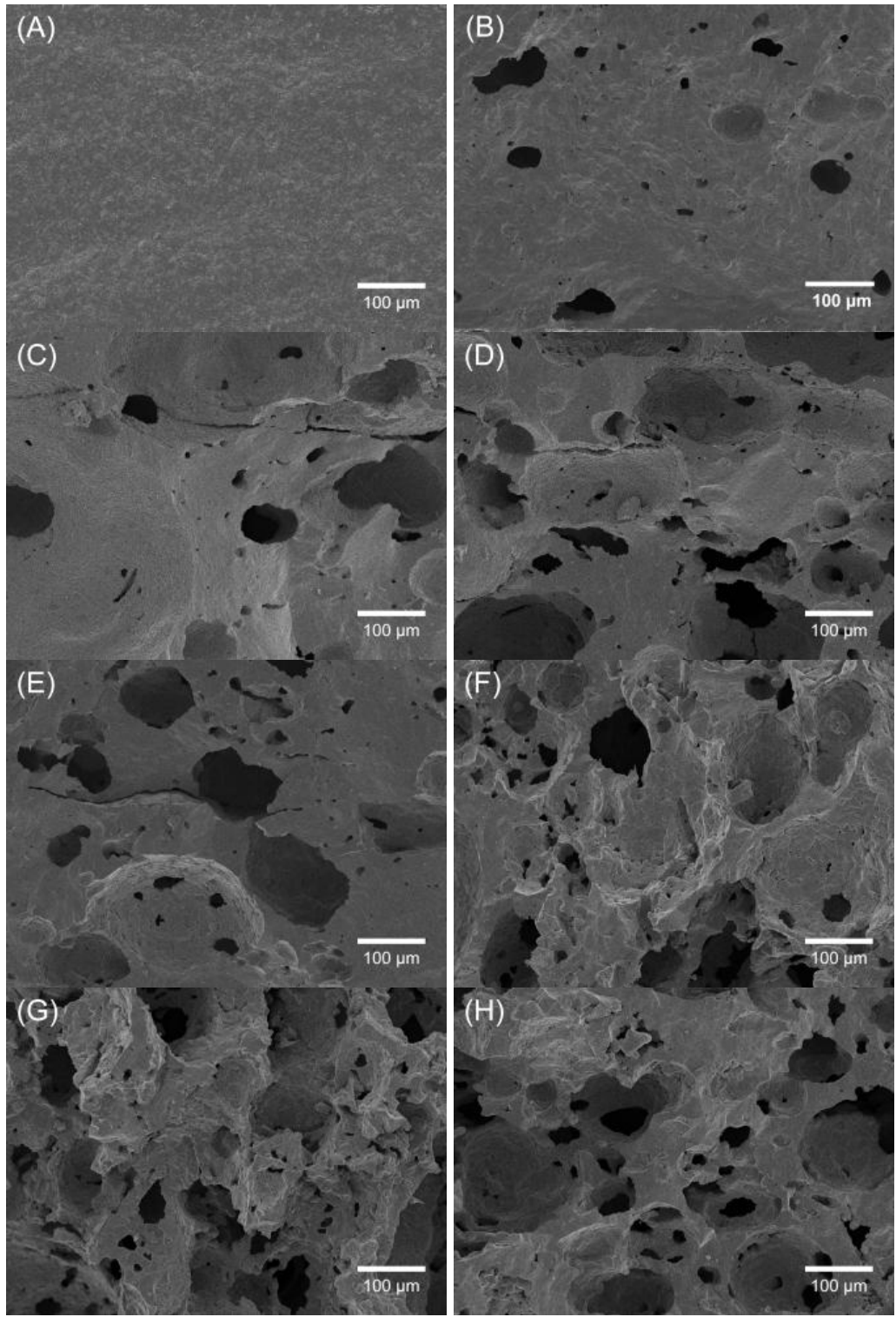

**Figure 1** Scanning electron micrographs of fracture surfaces of (A) dense (4 vol.%) and porous BCZT with the porosity of (B) 10 vol.%, (C) 15 vol.%, (D) 20 vol.%, (E) 25 vol.%, (F) 30 vol.%, (G) 35 vol.%, (H) 40 vol.%.

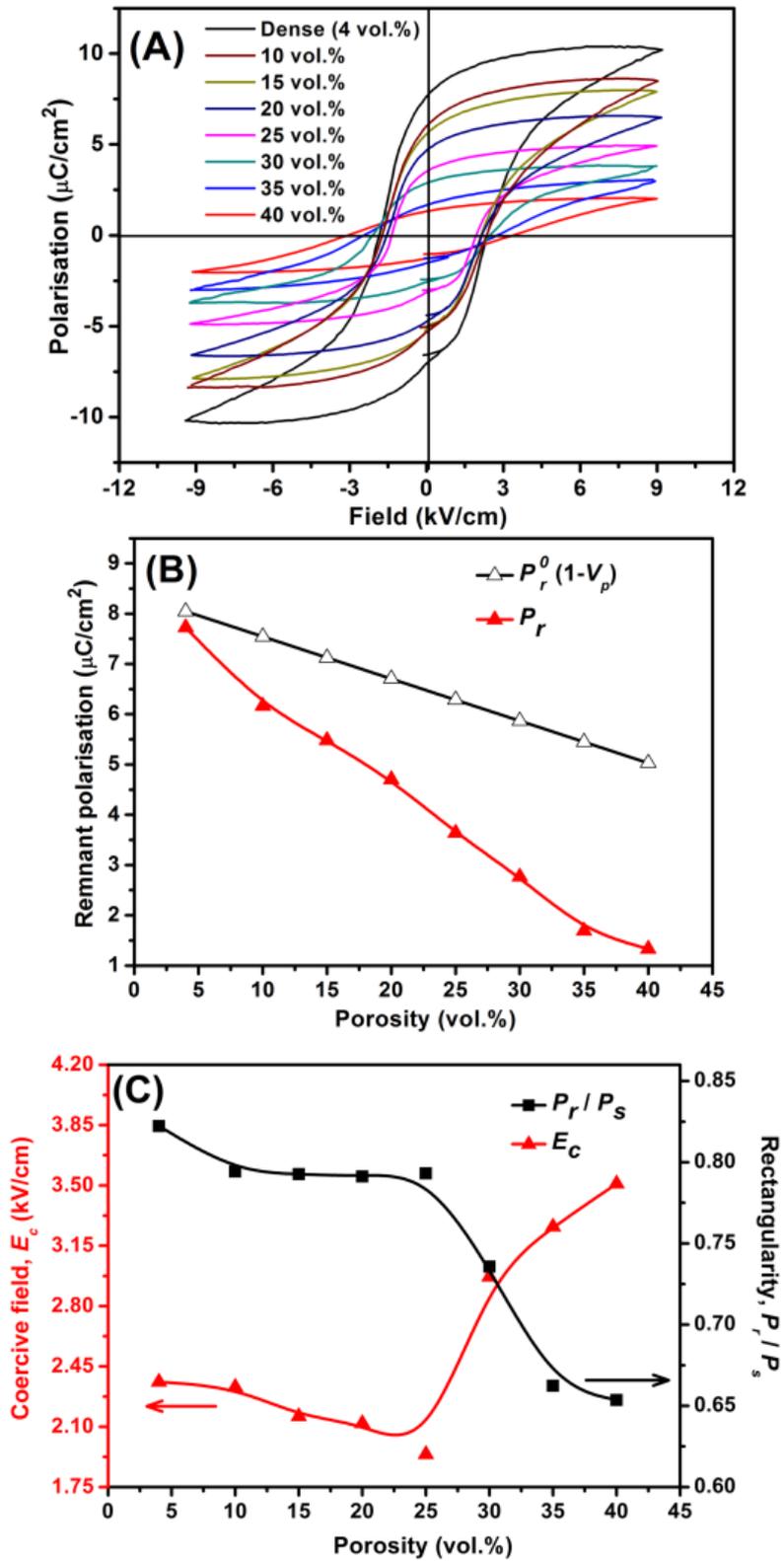

**Figure 2** (A) polarisation-field loop for range of $v_p$ with the hysteresis period of 200 ms, (B) remnant polarisation and (C) coercive field and rectangularity of BCZT with randomly distributed porosity; microstructures are in Figure 1.

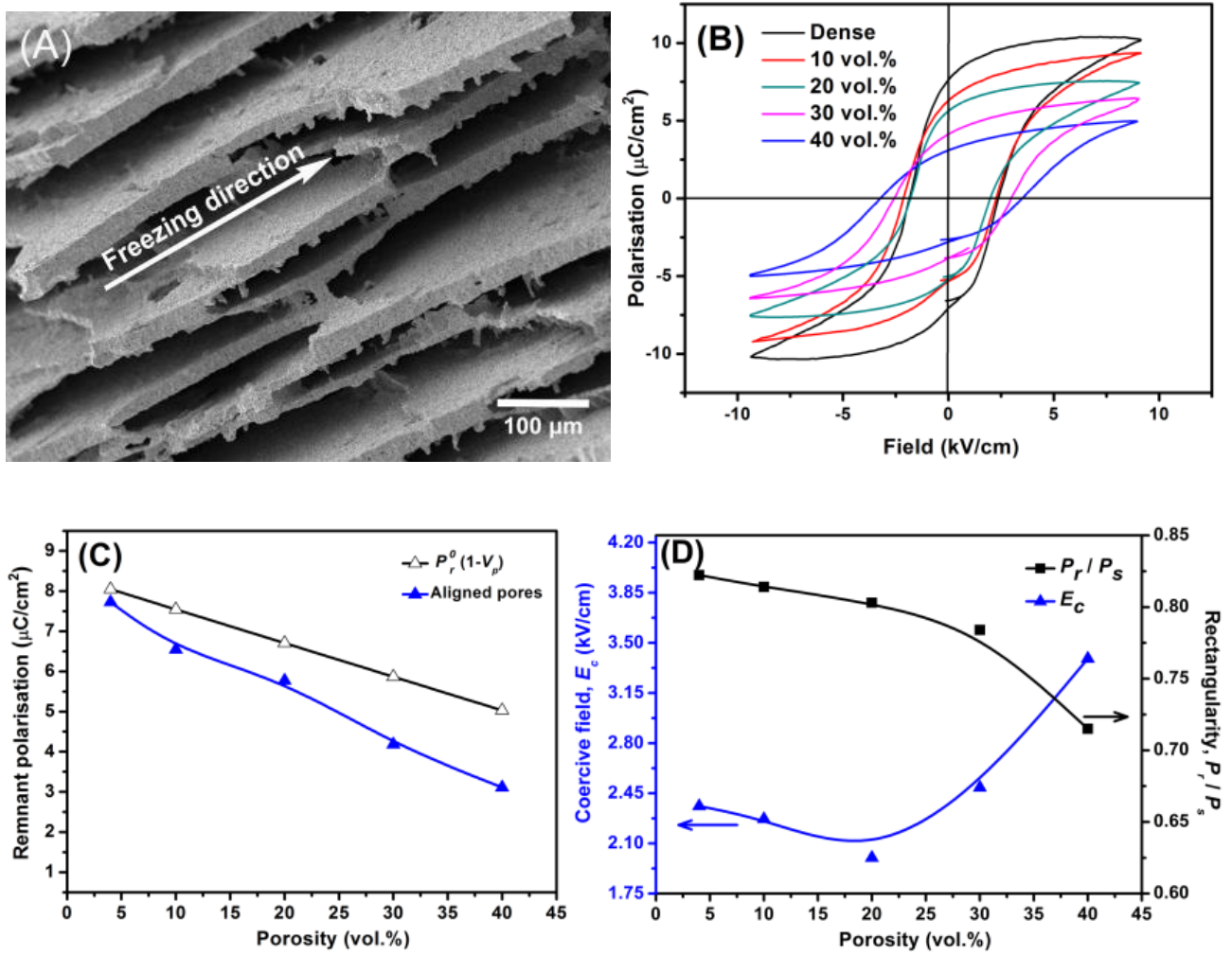

**Figure 3** (A) Scanning electron microscopy image of porous freeze-cast BCZT with porosity of 40 vol.% with the hysteresis period of 200 ms, (B) polarisation-field loops along freezing direction with different pore volume fractions, (C) remnant polarisation and (D) coercive field and rectangularity of freeze-cast BCZT.

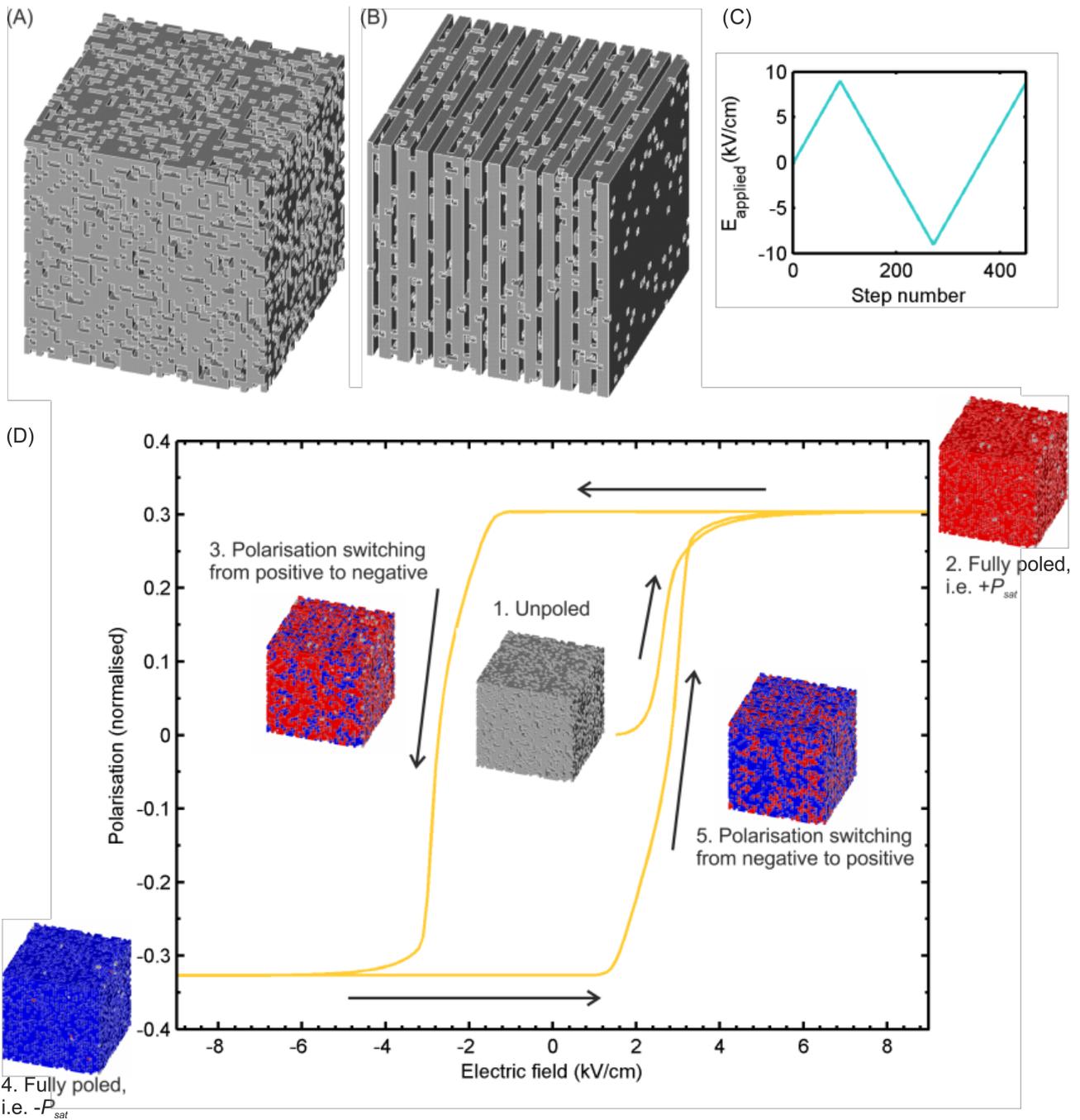

**Figure 4** Example model geometry for (A) equi-axed pores (30 vol.% porosity) and (B) aligned pores (33 vol.% porosity), (C) Voltage profile applied to the model to generate polarisation-field loops, and (D) is a schematic of the modelling process.

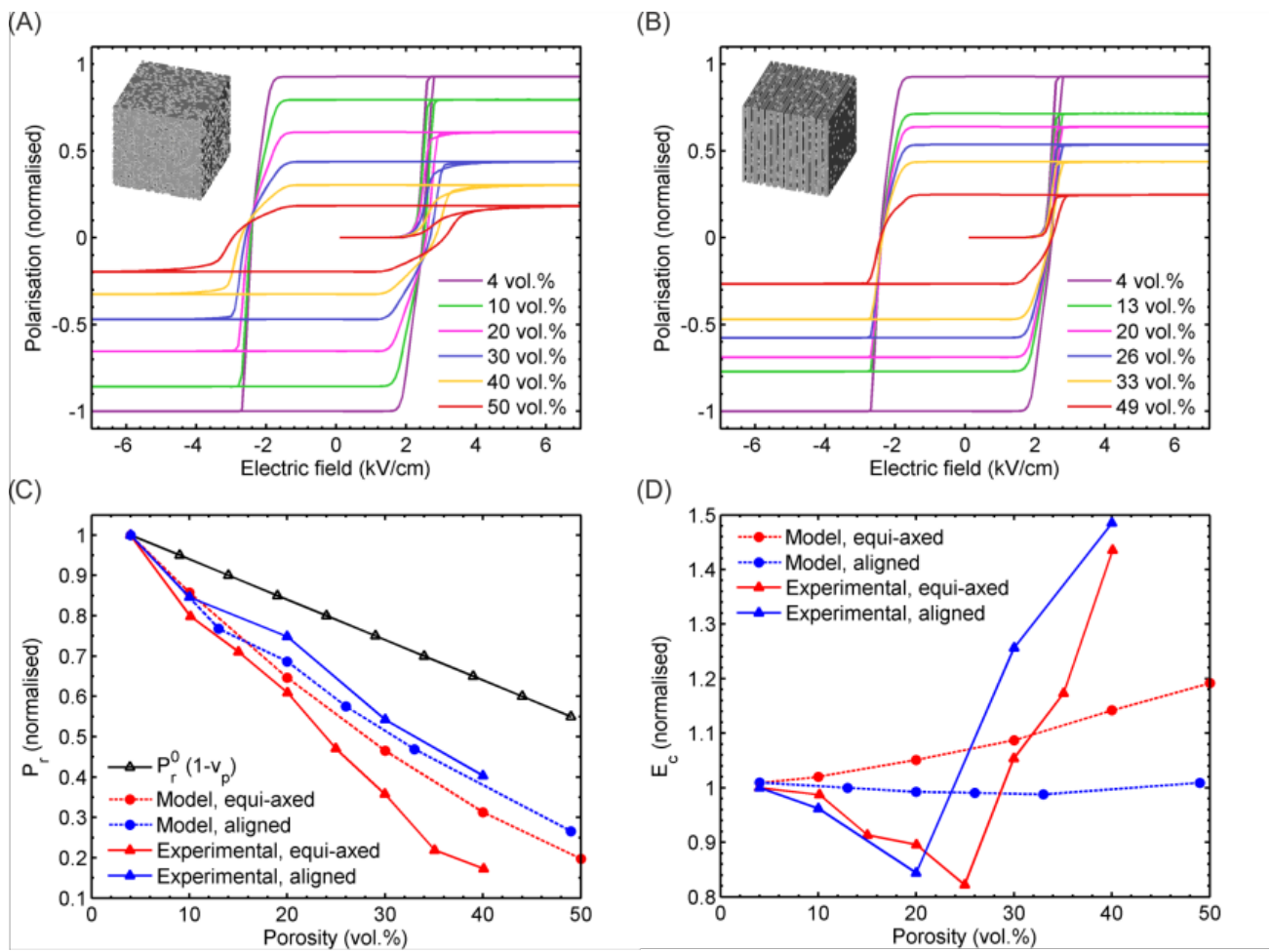

**Figure 5** Polarisation-field loops for (A) BCZT with random porosity and (B) for BCZT with aligned porosity, and variation of (C) remnant polarisation and (D) coercive field with porosity of both model and experimental data (normalised to the properties of the dense BCZT sample).

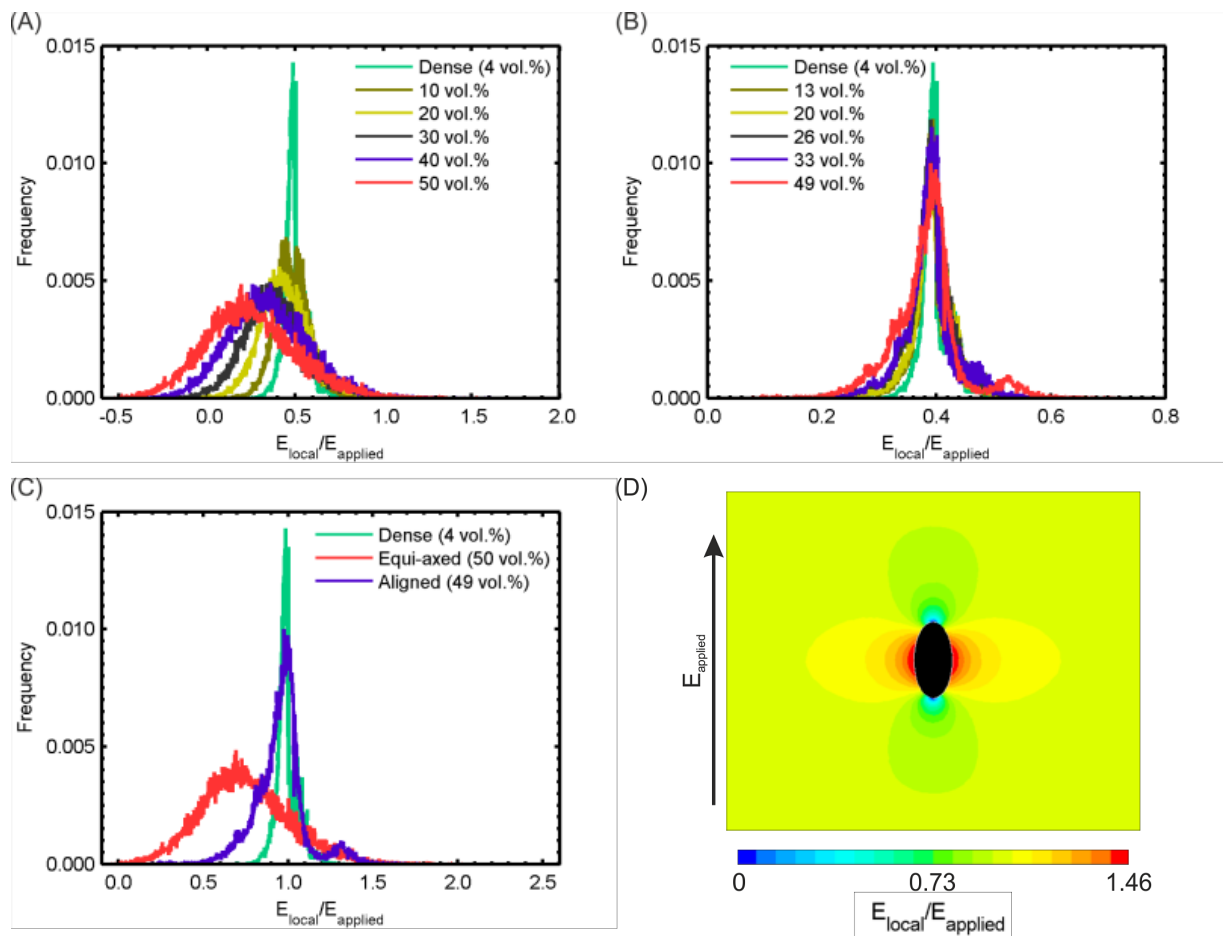

**Figure 6** Electric field distributions from modelling with (A) equi-axed porosity (vol.%) distributed at random in the structure, (B) aligned porosity (vol.%) aligned to the direction of the applied field, to simulate the type of structure obtained via the freeze casting process, and (C) a comparison of the spectra for dense BCZT (4 vol.% porosity) and BCZT with ~50 vol.% equi-axed and aligned porosities; field distribution curves have been normalised to the total number of ferroelectric elements in each model, such that the interval under each curve is one. (D) Electric field contour plot from two-dimensional finite element model demonstrating the inhomogeneous field distribution in the vicinity of a single pore.

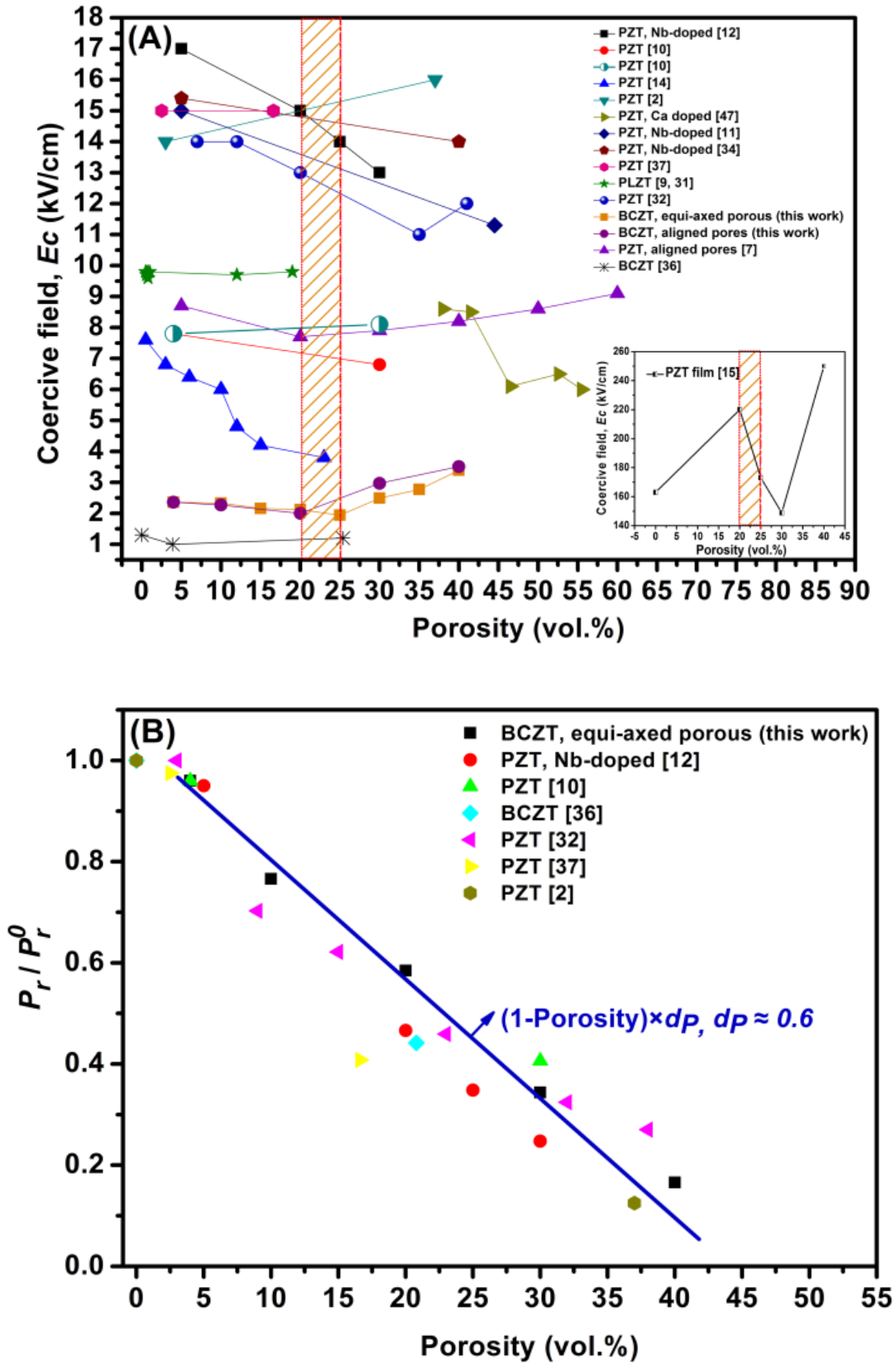

**Figure 7** (A) Change of coercive field $E_c$ with porosity (PZT film is plotted in inset as $E_c$ is much higher than other materials) and (B) effect of equi-axed porosity on remnant polarisation, $P_r$, normalised with respect to the dense material, $P_r^0$ [2, 7, 9-12, 14, 31, 32, 34, 36, 37, 47].